\documentclass[prl,a4paper,twocolumn,showpacs,superscriptaddress]{revtex4}
\usepackage{amsmath}
\usepackage{amsfonts}
\usepackage{amssymb}
\usepackage{bm}
\usepackage{graphicx}
\begin{document}
\title{Structure of surface vortex sheet between two rotating $^3$He superfluids}

\author{R. H\"anninen}
\affiliation{Low Temperature Laboratory, Helsinki University of Technology, P.O.Box
2200, FIN-02015 HUT, Finland }

\author{R.~Blaauwgeers}
\affiliation{Low Temperature Laboratory, Helsinki University of Technology, P.O.Box
2200, FIN-02015 HUT, Finland } \affiliation{Kamerlingh Onnes Laboratory, Leiden
University, 2300 RA Leiden, The Netherlands}

\author{V.~B.~Eltsov}
\affiliation{Low Temperature Laboratory, Helsinki University of Technology, P.O.Box
2200, FIN-02015 HUT, Finland } \affiliation{Kapitza Institute for Physical Problems,
Kosygina 2, 119334
  Moscow,  Russia}

\author{A.~P.~Finne}
\affiliation{Low Temperature Laboratory, Helsinki University of Technology, P.O.Box
2200, FIN-02015 HUT, Finland }

\author{M.~Krusius}
\affiliation{Low Temperature Laboratory, Helsinki University of Technology, P.O.Box
2200, FIN-02015 HUT, Finland }

\author{E. V. Thuneberg}
\affiliation{Low Temperature Laboratory, Helsinki University of Technology, P.O.Box
2200, FIN-02015 HUT, Finland } \affiliation{Department of Physical Sciences, P.O.Box
3000, FIN-90014 University of Oulu, Finland}

\author{G.~E.~Volovik}
\affiliation{Low Temperature Laboratory, Helsinki University of Technology, P.O.Box
2200, FIN-02015 HUT, Finland } \affiliation{Landau Institute for Theoretical
Physics, Kosygina 2, 119334
  Moscow, Russia}

\date{\today}
\begin{abstract}
We study a two-phase sample of superfluid $^3$He where vorticity exists in one phase
($^3$He-A) but cannot penetrate across the interfacial boundary to a second coherent
phase ($^3$He-B). We calculate the bending of the vorticity into a surface vortex
sheet on the interface and solve the internal structure of this new type of vortex
sheet. The compression of the vorticity from three to two dimensions enforces a
structure which is made up of ${\textstyle \frac {1} {2}}$-quantum units,
independently of the structure of the source vorticity in the bulk. These results
are consistent with our NMR measurements.
\end{abstract}
 \maketitle


\newcommand{\dvec}{{\hat {\bf d}}}
\newcommand{\lvec}{{\hat {\bf l}}}
\newcommand{\mvec}{{\hat {\bf m}}}
\newcommand{\nvec}{{\hat {\bf n}}}
\newcommand{\bnabla}{\mbox{\boldmath$\bm{\nabla}$}}
\newcommand{\vsA}{{\bf v}_{\rm sA}}
\newcommand{\vsB}{{\bf v}_{\rm sB}}
\newcommand{\vn}{{\bf v}_{\rm n}}

Consider an interface separating two phases whose order parameters are coherent
across the boundary. How do topological line or planar defects behave when they meet
the boundary? The coherence rules out simple termination at the interface. The
remaining alternatives are that the defect crosses the boundary or is deflected to
continue along the boundary. If deflected, how does the defect bend onto the
interface and what is its structure when it lies on the boundary? Such questions are
certainly relevant for dislocations at coherent grain boundaries in crystals. These
questions also appear in liquid crystals \cite{Kleman} and in the cosmos
\cite{KibbleClassification,MonopoleErasure}. Here we provide an answer in the
context of superfluid $^3$He, where detailed results can be achieved by combining
experimental and theoretical analysis \cite{MovingAB,KH-Instability}.

\begin{figure}[t]
\centerline{\includegraphics[width=0.8\linewidth]{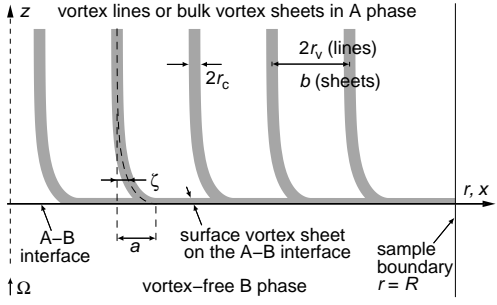}}
\medskip
\caption{A-phase vorticity (drawn in cross section) curves into a surface vortex
sheet on the A-B interface ($\Omega \sim 1\,$rad/s). } \label{LineBending}
\end{figure}

The important defects in superfluids are vortex lines or sheets, which can be
created in a controlled way by rotation. A crucial property of $^3$He superfluids
are multiple length scales: The core diameter of a typical vortex is $10^3$ times
larger in the A phase than in the B phase. Correspondingly the vortex energy in the
A phase is lower and vortices do not easily penetrate from the A to the B phase, but
form a {\it surface sheet} on the phase boundary (Fig.~\ref{LineBending}). Here we
calculate how the vorticity in bulk $^3$He-A bends to form such a surface sheet. The
calculated internal structure of this surface sheet turns out to be quite different
from the vortex sheet that appears in bulk $^3$He-A. Finally, we report on NMR
measurements and show that our calculations are consistent with these.

To obtain a two-phase sample of superfluid $^3$He, the A-B interface is stabilized
in a gradient of magnetic field \cite{KH-Instability}. Vortices are created by
rotating the sample at angular velocity $\Omega$ around the axis $\hat{\bm{z}}$
perpendicular to the interface. In the A-phase section vortices are formed at a low
critical velocity $v_{\rm cA}$ so that the average superfluid velocity $\langle
\bm{v}_{\rm sA}\rangle$ approximates solid-body rotation, $\langle \bm{v}_{\rm
sA}\rangle\approx\bm{\Omega}\times\bm{r}$. Depending on preparation, the vorticity
in the bulk is in the form of vortex lines or vortex sheets
\cite{VorSheetStructure,VorSheetDynamics}. Both structures have a large ``soft
vortex core" region for which the length scale $\xi_{\rm d} \sim 10\,\mu$m is set by
weak dipole-dipole forces. The vortex line is doubly quantized in units of the
circulation quantum $\kappa_0=h/2m$, where $m$ is the mass of a $^3$He atom. The
bulk vortex sheet has periodic units, which also consist of two quanta. The B phase
vortices, in turn, are singly quantized and have a narrow ``hard vortex core" with a
radius comparable to the superfluid coherence length $\xi \sim 10\,$nm. Their
smaller core radius causes the B-phase critical velocity to be at least an order of
magnitude larger. Thus the B phase remains in metastable vortex-free state, where
its superfluid fraction is stationary in the laboratory frame: $v_{\rm sB}=0$. To
sustain the difference in superflow velocities at the A-B interface, the A-phase
vorticity has to curve onto the interface where it forms a surface vortex sheet
(Fig.~\ref{LineBending}).

{\it Bending of vorticity into surface sheet:---}We  calculate the macroscopic
configuration for bending the vorticity, both in the form of vortex lines and bulk
sheets. For vortex lines we use the Bekarevich-Khalatnikov model \cite{Khalatnikov}.
The coarse-grained vorticity $\bm{\omega} = \frac {1} {2} \bm{\nabla}\times
\langle\bm{v}_{\rm s}\rangle$ is determined from the energy functional
\begin{equation}
 F= {\textstyle \frac {1}{2}} \rho_s\int d^3r\left[ \gamma|\bm{\omega}| +
\left( \langle \bm{v}_{\rm s}\rangle-\bm{v}_{\rm n}\right)^2\right]~.
\label{FreeEnergyRotation}
\end{equation}
Here $\rho_s$ is the density of the superfluid fraction. The first term is the
energy of vortex lines, where $\gamma = (2\kappa_0/\pi) \ln{(r_{\rm v}/r_{\rm c})}$
(Fig.~\ref{LineBending}) is on the order of the circulation of a line.  The second
term is the energy penalty associated with the difference between the average
superfluid and normal velocities, $\langle \bm{v}_{\rm s}\rangle$ and $\bm{v}_{\rm
n}=\bm{\Omega} \times\bm{r}$. For vortex sheets we use the functional
\begin{equation}
 F= {\textstyle \frac {1}{2}} \rho_s\int d^3r
\left( \bm{v}_{\rm s}-\bm{v}_{\rm n}\right)^2+\int_{\rm sheet}d^2r\,\sigma~.
\label{FreeEnergyVS}
\end{equation}
The latter term is the surface energy of the bulk sheet, with the surface tension
$\sigma$. In contrast to Eq. (\ref{FreeEnergyRotation}), the velocity $\bm{v}_{\rm
s}=(\hbar/2m)\bm{\nabla}\phi$ is calculated exactly ($\nabla^2\phi=0$), treating the
bending sheet as a tangential discontinuity of $\bm{v}_{\rm s}$.

In rotation with $\bm{\Omega}=\Omega\hat{\bm{z}}$, we assume the vorticity to occupy
the half space $z>0$ and not to penetrate to $z< 0$. We select a location far from
the axis of rotation so that the cylindrical coordinates can locally be approximated
with cartesian coordinates, taking $x$ in the radial direction. We assume that the
vortex sheets at $z\rightarrow \infty$ are perpendicular to $x$
\cite{VorSheetStructure}. You then find that between two sheets
$\bm{v}=\bm{v}_s-\bm{v}_n=2\Omega(x-c)\hat{\bm{y}}$, where $c$ is a constant. This
allows us to write Eq.\ (\ref{FreeEnergyVS}) as
\begin{equation}
F=\int _0^\infty dz\left[\frac{\rho_s \Omega^2}{6}(b^2+12\zeta^2)
+\frac{\sigma}{b}\sqrt{1+\left(\frac{d\zeta}{dz}\right)^2}\right]
\label{e.fvssimple}\end{equation} Here $b=(3\sigma/\rho_s\Omega^2)^{1/3}$ is the
separation of two sheets in the bulk ($z\rightarrow \infty$) and $\zeta$ their
radial displacement.

Surprisingly we find that both models [(\ref{FreeEnergyRotation}) and
(\ref{e.fvssimple})] give exactly the same form of bending: In spite of different
physics and approximations, the radial deviation $\zeta(z)$ of both vortex lines and
vortex sheets is given by
\begin{equation}
 \frac{z}{a}=1-\sqrt{2-(\zeta/a)^2}-
\frac{1}{\sqrt{2}}\ln\frac{\sqrt{2}-\sqrt{2-(\zeta/a)^2}}{(\sqrt{2}-1)\zeta/a}
\end{equation}
For vortex lines $a=\sqrt{\gamma/(4\Omega)}$ is on the order of the line spacing
$2r_{\rm v}$. For vortex sheets $a=b/\sqrt{6}$ is 0.41 times their equilibrium
spacing in the bulk. The bending contour plotted in Fig.~\ref{LineBending} becomes
horizontal at the radial deviation $\zeta =a$. After this point the bulk vorticity
is transformed to a new state of vortex matter -- a surface sheet. The vorticity in
the sheet grows linearly with distance $r$ from the center and finally escapes to
the vertical sample boundary.

{\it Structure of surface vortex sheet:---}The order parameter of superfluid $^3$He,
a $3\times 3$ matrix $A_{\mu j}$, takes in the A phase the form $A_{\mu
j}=\Delta_{\rm A}\hat{d}_\mu(\hat{m}_j+{\rm i}\hat{n}_j)$. Here $\hat{\bm{m}}$,
$\hat{\bm{n}}$, and $\hat{\bm{l}}$ are three orthogonal unit vectors that specify
the orbital part of the Cooper pair wave function. The vector $\hat{\bm{d}}$
specifies the orientation of the spin part. Here we consider only continuous vortex
structures where the amplitude $\Delta_{\rm A}$ is constant, and the circulation
arises solely from a smooth orientational winding of the triad ($\hat{\bm{m}}$,
$\hat{\bm{n}}$, $\hat{\bm{l}}$) in space. The vortex textures are determined from
the energy functional $F=\int_{\rm A} d^3r\,f_{\rm A}+ \int_{\rm B} d^3r\, f_{\rm
B}$, where \cite{VW}
\begin{eqnarray} &2f_{\rm A}&=
\rho_\perp\bm{ v}_{\rm A}^2+(\rho_\parallel-\rho_\perp)(\hat{\bm{l}}\cdot\bm{v}_{\rm
A})^2 +2C\bm{ v}_{\rm A}\cdot\bm{\nabla}\times\hat{\bm{l}} \nonumber \\&&
-2C_0(\hat{\bm{l}}\cdot\bm{v}_{\rm A})
(\hat{\bm{l}}\cdot\bm{\nabla}\times\hat{\bm{l}})+ K_{\rm
s}(\bm{\nabla}\cdot\hat{\bm{l}})^2 \nonumber \\&&+K_{\rm
t}(\hat{\bm{l}}\cdot\bm{\nabla}\times\hat{\bm{l}})^2+K_{\rm
b}\vert\hat{\bm{l}}\times(\bm{\nabla}\times\hat{\bm{l}})\vert^2 \nonumber \\ &&+K_5
\vert(\hat{\bm{l}}\cdot\bm{\nabla})\hat{\bm{d}}\vert^2+
K_6\sum_{ij}[(\hat{\bm{l}}\times\bm{\nabla})_i\hat{\bm{d}}_j)]^2\nonumber
\\&& +\lambda_{\rm d}\vert\hat{\bm{d}}\times\hat{\bm{l}}\vert^2 +\lambda_{\rm
h}(\hat{\bm{d}}\cdot\bm{H})^2, \label{e.hydrof} \end{eqnarray} $\bm{v}_{\rm
A}=\bm{v}_{\rm sA}-\bm{v}_{\rm n}$ and $\bm{v}_{\rm sA}=(\hbar/2m)\sum_i\hat
m_i\bm{\nabla}\hat n_i$. The first nine terms give the gradient energy, while the
last two terms are the dipole-dipole and external field energies. This functional
replaces Eqs.\ (\ref{FreeEnergyRotation}) and (\ref{FreeEnergyVS}) when the
resolution is increased from $r_{\rm v}$ to the dipolar coherence length $\xi_{\rm
d}=(\hbar/2m)\sqrt{\rho_\parallel/\lambda_{\rm d}}$. At the A-B interface one has
\cite{BoundaryCondition}
\begin{equation}
\hat{\bm{d}}=   \tensor{R} \cdot \hat{\bm{s}}\,, \hspace{10mm} (\hat{\bm{m}}+ {\rm
i} \hat{\bm{n}}) \cdot \hat{\bm{s}} =   e^{{\rm i} \phi}\,, \hspace{10mm}
\hat{\bm{l}}\cdot \hat{\bm{s}} = 0 \label{e.lbound}
\end{equation}
as boundary conditions, where $\hat{\bm{s}}$ is the normal of the interface. The
rotation matrix $\tensor{R}$ and the phase angle $\phi$ are quantities appearing in
the B-phase order parameter $A_{\mu j}=\Delta_{\rm B}e^{{\rm i}\phi} R_{\mu j}$. On
the A-phase side the role of the phase angle is played by the rotation angle of
$\hat{\bm{m}}$ and $\hat{\bm{n}}$ around $\hat{\bm{l}}$. The boundary conditions
(\ref{e.lbound}) imply the coherence of the phase angle across the interface.

We calculate the order parameter in the surface sheet by minimizing the total energy
$F$ numerically.  The main assumption is that the solutions are homogeneous in the
(radial) $x$ direction and periodic in the perpendicular (azimuthal) $y$ direction.
Locally these assumptions are approximately satisfied everywhere except near points
where new vorticity enters the sheet. The velocities far above and below the sheet
satisfy $|v_{\rm B}^\infty - v_{\rm A}^\infty| = 2\kappa_0/L_{y}$ (or $L_y \! = \! 2
\kappa_0 /(\Omega r)$) when there are two circulation quanta per period $L_y$. Since
the velocity is effectively screened on the A-phase side by vorticity, we can take
$v_{\rm A}^\infty\approx 0$. The experiment is performed in a magnetic field
$\bm{H}\parallel \hat {\bm{z}}$ which locks $\hat{\bm{d}}\perp\bm{H}$ \cite{approx}.
Since the variation of $\hat{\bm{d}}$ has only a minor effect, we approximate the B
phase $\tensor{R}$ with a constant. In contrast, the phase field $\phi(y,z)$ on the
B phase side has to be properly included, via  $f_{\rm B}=\frac{1}{2}\rho_{\rm
s}(\bm{v}_{\rm sB}-\bm{v}_{\rm n})^2$ and $\bm{v}_{\rm sB}
=(\hbar/2m)\bm{\nabla}\phi$. Otherwise a surface vortex sheet is not stable, as is
the case for a sheet which would coat a solid container wall \cite{MovingAB}.

Depending on the density of vorticity in the surface sheet, we obtain two different
textures which both incorporate $\frac {1}{2}$-quantum vortex units. These textures
are independent of the initial ansatz, {\it ie.} whether a $2$-quantum vortex line
or one period of the bulk vortex-sheet is placed above the A-B interface at the
start of the iterative energy minimization. This means that the bulk structures
loose their identity and transform at the A-B interface to surface sheet textures.

\begin{figure}[t!!]
\includegraphics[width=8.5cm]{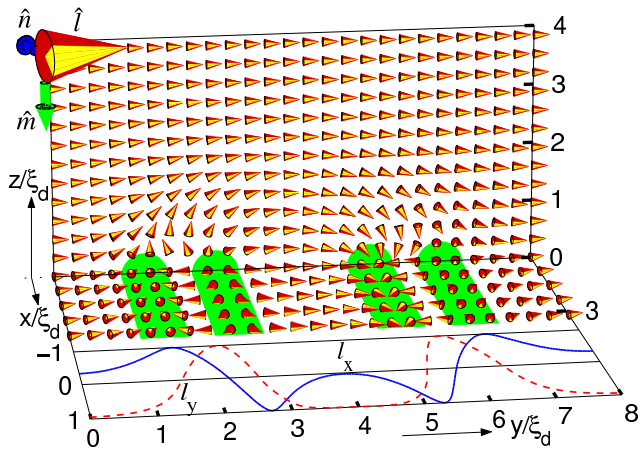}
\caption{Orbital texture of the surface sheet at low density of vorticity ($L_y =
8\, \xi_{\rm d}$) in a frame where $v_{\rm n}=0$. The cones point along
$\hat{\bm{l}}$ and their yellow stripes indicate the rotation of $\hat{\bm{m}}$ and
$\hat{\bm{n}}$ around $\hat{\bm{l}}$. On the A-B plane ($x-y$ plane) $\hat{\bm{l}}$
is parallel to the interface (Eq.~(\ref{e.lbound})). Its orientations there are
shown in the diagram on the bottom. The four highlighted regions ($\hat{\bm{l}}
\approx \pm \, \hat {\bm{x}}$) are the centers of ${\textstyle \frac {1}
{2}}$-quantum cores which pairwise form $1$-quantum composites. Two ${\textstyle
\frac {1} {2}}$-quantum cores in one pair are separated by $d_{\rm c} =$ $ 0.26 \,
\xi_{\rm d} + 0.135\, L_y - 0.0027\, L_y^2/\xi_{\rm d}$, when $ 5.5\, \xi_{\rm d}
\leq L_y$ $ \leq 20\, \xi_{\rm d}$.  } \label{f.u2d2}
\end{figure}

\begin{figure}[t!!]
\includegraphics[width=8.5cm]{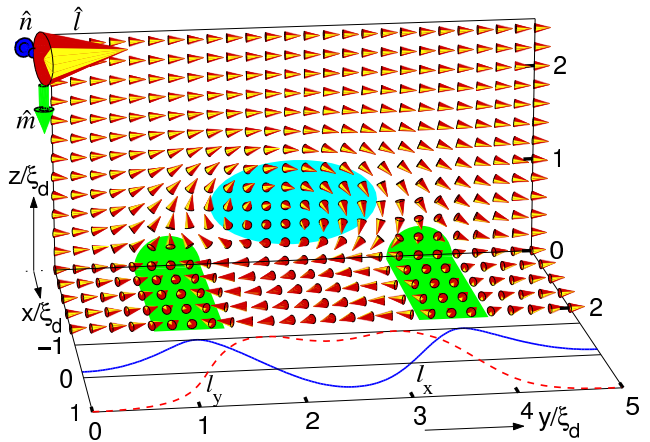}
\caption{Orbital texture of the surface sheet at high density of vorticity ($L_ y =
5\xi_{\rm d}$). With increasing density, the two ${\textstyle \frac {1}
{2}}$-quantum cores in the center of Fig.~\protect\ref{f.u2d2} form a $1$-quantum
composite, a circular {\it meron}, which pops above the A-B plane, as seen in this
figure. The highlighted regions on the A-B plane, the two outermost ${\textstyle
\frac {1} {2}}$-quantum vortices from Fig.~\protect\ref{f.u2d2}, are now isolated,
but weakly bound by the meron above the A-B interface. The separation between the
${\textstyle \frac {1} {2}}$-quantum cores is $d_{\rm c} = 0.0743\, \xi_{\rm d} +$
$0.507 \, L_y$, when $ 2 \xi_{\rm d} \leq L_y \leq 7.5 \xi_{\rm d}$. } \label{f.d2}
\end{figure}

The low-density texture in Fig.~\ref{f.u2d2} has all the vorticity aligned on the
A-B interface.
It separates into two composite 1-quantum vortices. These consist of two
${\textstyle \frac {1} {2}}$-quantum cores, although the vorticity $\bm{\nabla}
\times \bm{v}_{\rm sA} $ (not shown) cannot be divided into distinct ${\textstyle
\frac {1} {2}}$-quantum units. At high density a different packing of the vorticity
becomes energetically favorable. The resulting texture in Fig.~\ref{f.d2} has two
${\textstyle \frac {1} {2}}$-quantum cores on the A-B boundary and the remaining
circulation localized as a $1$-quantum vortex above the A-B interface.
Here the vorticity $\bm{\nabla} \times \bm{v}_{\rm sA} $ is maximized in the $\frac
{1}{2}$-quantum regions. The distinguishing feature of the different vortex
components is the solid angle which their orientational distribution of
$\hat{\bm{l}}$ covers. For instance, the circular $1$-quantum vortex above the A-B
plane in Fig.~\ref{f.d2} includes all orientations of the positive hemisphere with
respect to the $x$ axis, while the two $\frac {1}{2}$-quantum vortices each cover
one quadrant on the negative hemisphere.

The first-order transition between the two textures takes place at $L_y = 5.7 \,
\xi_{\rm d}$ or when the velocity difference in shear flow $|v_{\rm B}^\infty-v_{\rm
A}^\infty | \approx 2.8$ mm/s. This value is in the middle of our measuring range.
In the calculations the transition is hysteretic, especially on moving from high to
low density. Transitions between the two textures are thus expected as a function of
$\Omega$ and $r$: The density of the surface vorticity increases with $r$ as $\frac
{1}{2} \Omega r/\kappa_0$. Thus at high $\Omega$ one expects to find the low-density
texture in the center and the high-density one outside a critical radius.

\begin{figure}[t!!!]
\centerline{\includegraphics[width=1.0\linewidth]{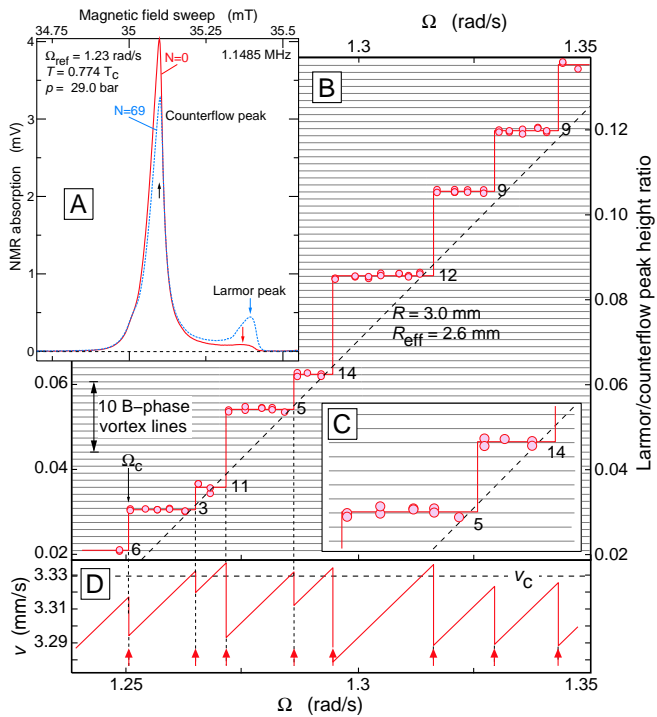}}
\medskip
\caption{NMR measurements on the shear-flow instability of the A-B interface provide
information on the A-phase surface vortex sheet. {\bf (A)} NMR absorption spectra of
$^3$He-B in the vortex-free state ($N=0$) and with 69 vortex lines ($N=69)$. The
ratio of the Larmor and counterflow peak heights, measured at constant conditions
(at $\Omega = \Omega_{\rm ref} < \Omega_{\rm c}$), is a linear function of $N$. {\bf
(B)} A repetitive sequence of instability events with increasing $\Omega$. The data
points plot the peak height ratio from NMR spectra measured at $\Omega_{\rm ref}$.
The spacing of the horizontal grid lines is the calibrated equivalent of one vortex
and yields the number of new B-phase vortex lines $\Delta N$ created in each
instability event (explicitly given at each discontinuous step). The dashed vertical
lines (and arrows in panel {\bf D}) denote the $\Omega$ values where the Larmor peak
height rises abruptly while $\Omega$ is slowly increased by a small increment. The
sloping dashed line is a fit through the corner points and defines the effective
radial location $R_{\rm eff}$ of the instability: $N = (2\pi/\kappa_0)\, R^2_{\rm
eff} \, (\Omega - \Omega_{\rm c})$. {\bf (C)} One of the instability events in
greater detail. The resolution allows us to conclude that $\Delta N$ is a small
random number, which can be even or odd. {\bf (D)} Discontinuous superflow velocity
with increasing $\Omega$ at $r=R_{\rm eff}$: $v=$ $|v_{\rm sB} - v_{\rm n}|= \Omega
R_{\rm eff} - \kappa_0 N/(2\pi R_{\rm eff})$. The horizontal dashed line is
equivalent to the sloping dashed line in panel {\bf B} and defines the mean critical
velocity $v_{\rm c} $. } \label{Steps+Flow+Config}
\end{figure}

{\it Transfer of vorticity across A-B interface:---}Our NMR measurements of the
two-phase sample show that the A-B interface is stable up to high rotation.
Ultimately at a critical angular velocity $\Omega_{\rm c}$, which corresponds to a
critical B-phase superflow velocity $v_{\rm c} \approx \Omega_{\rm c} R$ with
respect to the container wall at $r \approx R$, the A-B interface undergoes an
instability and a small number of circulation quanta manage to cross the interface
and form the first vortex lines in the B phase. An analysis of this event as a
function of temperature and barrier field shows that $v_{\rm c}$ is determined by
the hydrodynamic rigidity of the interface, on which the texture of the surface
vortex sheet has little effect. The upper limit for the density of vorticity in the
surface sheet is thus placed by the shear-flow instability. With our solenoidal
barrier magnet this means that $\Omega_{\rm c} \lesssim 1.6\,$rad/s. If the rotation
is increased above $\Omega_{\rm c}$, then the instability occurs repetitively at the
constant critical velocity $v_{\rm c}$, as analyzed in Fig.~\ref{Steps+Flow+Config}.

From Fig.~\ref{Steps+Flow+Config} it is concluded that the number of circulation
quanta $\Delta N$, which is transferred across the A-B interface in one event,  can
be either odd or even. Measurements accumulated in the constant conditions of
Fig.~\ref{Steps+Flow+Config} on more than 200 instability events display a smooth
probability density distribution $P(\Delta N)$, which is centered around $\langle
\Delta N \rangle \approx 8$ and where both odd and even values of $\Delta N$ are
equally likely. This indicates that the vorticity from the bulk A-phase breaks
through the A-B interface as single quanta, although the source object, the A-phase
vortex, is doubly quantized. Thus after an instability event the number of quanta in
the B-phase section of the sample can be odd, while it is in the bulk A phase even.
This means that one unpaired quantum must be accommodated and stored in the surface
sheet in stable state. These properties remain unchanged if the A-phase section of
the sample is arranged to support the bulk vortex sheet, in any of its different
global configurations, as explained in Ref.~\cite{VorSheetDynamics}. The
measurements support the conclusion from the texture calculations that, compared to
the bulk A-phase vortex textures, the surface sheet is made up from smaller building
blocks and has autonomous structure, independently of the bulk vortex textures above
the A-B interface. The experiments cannot as yet distinguish between the two
calculated textures, but in both cases the vorticity can be combined to $1$-quantum
units which allows the transfer of vorticity from the A to the B phase.

{\it Discussion:---}Stable vortex sheets are discussed in coherent quantum systems
with a multi-component order parameter. So far two examples have been experimentally
identified, both in anisotropic superfluid $^3$He-A, namely the bulk vortex sheet
\cite{VorSheetStructure} and the present surface sheet. The stability of their
structures is based on different principles. The bulk sheet is topologically stable:
Its linear quantized vorticity consists of $1$-quantum building blocks, known as
merons or Mermin-Ho vortices, which come pairwise as a combination of a circular and
a hyperbolic unit. These are confined within a domain-wall-like planar defect. In
the surface sheet the quantized vorticity is confined to two dimensions by the
hydrodynamic binding of the circulation to the interfacial boundary. The texture is
made up of $\frac {1}{2}$-quantum building blocks to satisfy the in-plane boundary
condition of the orbital anisotropy axis $\hat{\bm{l}}$ at the A-B interface
[Eq.~(\ref{e.lbound})]. The two calculated textures are the first with experimental
basis, where fractionally quantized units appear in $^3$He superfluids, since so far
the predicted isolated $\frac {1}{2}$-quantum vortex \cite{Half-QuantVor} has not
been observed in $^3$He-A.

This collaboration was carried out under the EU-IHP ULTI-3, the ESF COSLAB, and ESF
VORTEX programs.



\vspace{-5mm}
\begin{thebibliography}{99}

\vspace{-3mm}
\bibitem{Kleman}M. Kleman, O.D. Lavrentovich, {\it Soft Matter Physics: An
Introduction} (Springer-Verlag, Berlin, 2003).


\bibitem{KibbleClassification}  T.W.B. Kibble, in {\it Topological
Defects and Non-Equilibrium Dynamics of Symmetry Breaking Phase Tranitions}, eds.
Yu.M. Bunkov, H. Godfrin (Kluwer Academic Publ., Dodrecht, 2000), p. 7.


\bibitem{MonopoleErasure} L. Pogosian, T. Vachaspati,
 Phys. Rev. {\bf D~62}, 105005 (2000); Phys. Rev. Lett. {\bf 80}, 2281 (1998).


\bibitem{MovingAB} \"U. Parts {\it et al.}, Phys. Rev. Lett. {\bf 71}, 2951 (1993);
Physica B {\bf 197}, 376 (1994).


\bibitem{KH-Instability} R. Blaauwgeers {\it et al.}, Phys. Rev. Lett. {\bf 89}, 155301
(2002); Physica B, in print (2003).


\bibitem{VorSheetStructure} \"U. Parts {\it et al.}, Phys. Rev. Lett. {\bf 72}, 3839 (1994);
Physica B {\bf 210}, 311 (1995).


\bibitem{VorSheetDynamics} V.B. Eltsov {\it et al.}, Phys. Rev. Lett. {\bf 88}, 65301
(2002).


\bibitem{Khalatnikov} I.M. Khalatnikov, {\it An Introduction to the
Theory of Superfluidity} (Benjamin Inc., New York, 1965), p. 93.


\bibitem{VW} D. Vollhardt and P. W\"olfle, {\it The superfluid phases of helium 3}
(Taylor \& Francis, London 1990).


\bibitem{BoundaryCondition} M.C. Cross, in {\it Quantum Fluids and Solids},
eds. S.B. Trickey, E.D. Adams, J.W. Duffy (Plenum Press, New York, 1977), p. 183.


\bibitem{approx} At $z \rightarrow +\infty$ outside
the soft core regions $\hat{\bm{l}}$ is homogenous. This leads to
$\hat{\bm{l}}(z=+\infty)\parallel\hat{\bm{y}} $. In addition, the coefficients of
the functional (\ref{e.hydrof}) are evaluated in the weak-coupling approximation at
temperatures $T\approx T_{\rm c}$.

\bibitem{Half-QuantVor} M. Salomaa, G. Volovik, Rev. Mod. Phys. {\bf 59}, 533 (1987).

\end{thebibliography}
\end{document}